\documentclass[12pt]{article}
\usepackage{epsf}
\usepackage{amsmath}
\usepackage{graphics}
\usepackage{cite}

\renewcommand{\thefootnote}{\#\arabic{footnote}}
\begin{document}

\newcommand{\gtrsim}{ \mathop{}_{\textstyle \sim}^{\textstyle >} }
\newcommand{\lesssim}{ \mathop{}_{\textstyle \sim}^{\textstyle <} }

\newcommand{\rem}[1]{{\bf #1}}

\renewcommand{\thefootnote}{\fnsymbol{footnote}}
\setcounter{footnote}{0}
\begin{titlepage}

\def\thefootnote{\fnsymbol{footnote}}

\begin{center}
\hfill November 2015\\
\vskip .5in
\bigskip
\bigskip
{\Large \bf The Primordial Black Hole Mass Range}

\vskip .45in

{\bf Paul H. Frampton\footnote{email: paul.h.frampton@gmail.com\\
homepage:www.paulframpton.org}}

{\em 4 Lucerne Road, Oxford OX2 7QB, UK.}

\end{center}

\vskip .4in
\begin{abstract}
\noindent
We investigate Primoridal Black Hole (PBH) formation by which we mean
black holes produced in the early universe during 
radiation domination.
After discussing the range of PBH mass permitted in the original
mechanism of Carr and Hawking, hybrid inflation
with parametric resonance is presented as an existence theorem for 
PBHs of arbitrary mass. As proposed in arXiv:1510.00400,
PBHs with many solar masses can provide
a solution to the dark matter problem in galaxies. 
PBHs can also explain dark matter observed
in clusters and suggest a primordial origin for supermassive 
black holes in galactic cores.
\end{abstract}

\end{titlepage}

\renewcommand{\thepage}{\arabic{page}}
\setcounter{page}{1}
\renewcommand{\thefootnote}{\#\arabic{footnote}}

\newpage

\section{Introduction}

\bigskip

\noindent
According to global analyses of the cosmological parameters
one quarter, or slightly more, of the energy of the universe is in the form of dark matter
whose constituent is the subject of the present paper. Recently it has been proposed
\cite{Frampton} that the dark matter constituents are black holes with masses many
times the mass of the Sun. In a galaxy like the Milky Way, the proposal is that
residing in the galaxy are between ten million and ten billion black holes with masses between
one hundred and one hundred thousand solar masses.

\bigskip

\noindent
Black holes in this range of masses are commonly known as Intermediate Mass Black Hole (IMBHs)
since they lie above the masses of stellar-mass black holes
and below the masses of the
supermassive black holes. It has long been mysterious
why there is a mass gap between stellar-mass and supermassive black holes.
If the proposed solution of the dark matter problem is correct, it will answer this old
question.

\bigskip

\noindent
There is irrefutable evidence for stellar-mass black holes from observations of X-ray binaries.
Such systems were first emphasized in \cite{Guseynov} then further studied in \cite{Trimble}. All
the known stellar-mass black holes are members of X-ray binaries. The first was
discovered over fifty years
ago in 1964 in Cygnus X-1 and many stellar-mass black holes have since been discovered
from studies of X-ray binaries, with masses in a range between 
$5M_{\odot}$ and  $100M_{\odot}$,
where the first-discovered Cygnus X-1 is at about $15M_{\odot}$.

\bigskip

\noindent
There is irrefutable observational evidence also for supermassive black holes from the
observations of fast-moving stars around them and such stars being swallowed or 
torn apart by the strong gravitational field. The first discovered SMBH was naturally
the one, Sag $A^*$, at the core of the Milky Way which was discovered
in 1974 and has mass $M_{SagA*} \sim 4.1 \times 10^6M_{\odot}$.
SMBHs discovered at galactic cores include those for galaxies named M31,
NGC4889, among many others. The SMBH at the core of 
the nearby Andromeda galaxy ($M31$)
has mass $M=2\times 10^8M_{\odot}$, fifty times $M_{SagA*}$.
The most massive core SMBH so far observed is for NGC4889 with 
$M \sim 2.1\times 10^9M_{\odot}$. Some galaxies contain two SMBHs in a binary, believed to be the result of a galaxy merger. 
Quasars contain black holes with even higher masses up to at least
$4\times 10^{10}M_{\odot}$. 

\bigskip

\noindent
We note historically that dark matter was first discovered by Fritz
Zwicky\cite{Zwicky,Zwicky2} in 1933
in the Coma Cluster, and its presence in galaxies was demonstrated
convincingly by Vera Rubin in the 1960s and 1970s 
from the rotation curves of many galaxies 
\cite{Rubin}.  Rubin has more recently made a prescient remark 
about not liking 
a universe filled with a new kind of elementary particle and 
we shall return to this, 
with the full quote, at the end of our final discussion.

\bigskip

\noindent
Regarding the PBH mass range, the purpose of the present article is to convince
the reader that the possible PBH masses extend upwards to many solar
masses and above, far beyond
what was was thought possible not many years ago when ignorance about
PBHs with many solar masses probably prevented the MACHO
\cite{MACHO} and EROS\cite{EROS} Collaborations from 
discovering all the dark matter.

\bigskip

\noindent
The plan of the present paper is that in Section 2 we review the original
implementation {\it \`{a} la} Carr and Hawking of PBH formation. In Section 3 we shall 
discuss parametric resonance in hybrid inflation which can produce
PBHs with arbitrary mass. In Section 4 possible implications
are discussed especially for 
dark matter but also for galactic-core supermassive black holes and
unassociated black holes. In Section 5
there is some final discussion.

\bigskip

\section{PBHs \`{a} la Carr and Hawking}

\bigskip
\noindent
If all black holes were formed by gravitational collapse then black holes with
$M_{BH} \ll M_{\odot}$ would be impossible because stars powered by nuclear fusion
cannot be far below $M = M_{\odot}$.
It was first suggested by Zel'dovich\cite{Zeldovich1,Zeldovich2} and by Hawking\cite{Hawking2}
that black holes can be produced in the early stages of the cosmological expansion\cite{Carr}.

\bigskip

\noindent
Such PBHs are of special interest for several reasons. Firstly, they are the only type
of black hole which can be so light, down to $10^{12}kg\sim10^{-18}M_{\odot}$, that Hawking radiation might conceivably be detected
\footnote{We shall, however, confirm at the end of this Section that such detection
is impracticable.}. Secondly, PBHs in the intermediate-mass region
$100M_{\odot} \leq M_{IMBH} \leq 10^6M_{\odot}$ can provide the galactic
dark matter. Thirdly, supermassive PBHs with $M_{SMBH} \geq 10^6M_{\odot}$ 
can play a role at galactic centers and provide some of the cluster dark matter.

\bigskip

\noindent
The mechanism of PBH formation involves large fluctuations or inhomogeneities. 
Carr and Hawking\cite{CH}
argued that we know there are fluctuations in the universe in order to seed
structure formation and there must similarly be fluctuations in the early universe. Provided the
radiation is compressed to a high enough density, meaning to a radius as small
as its Schwarzschild radius, a PBH will form. Because the density in the early
universe is extremely high, it is very likely that PBHs will be created. The two
necessities are high density which is guaranteed and large inhomogeneities.

\bigskip

\noindent
During radiation domination
\begin{equation}
a(t) \propto t^{1/2}
\label{raddom}
\end{equation}
and 
\begin{equation}
\rho_{\gamma} \propto a(t)^{-4} \propto t^{-2}
\label{rhogamma}
\end{equation}

\bigskip

\noindent
Ignoring factors $O(1)$, as we shall do throughout this paper, and bearing in
mind that the radius of a black hole is 
\begin{equation}
r_{BH} \sim \left( \frac{M_{BH}}{M_{Planck}^2} \right)
\label{BHradius}
\end{equation}
with
\begin{equation}
M_{Planck} \sim 10^{19}GeV \sim 10^{-8} kg \sim 10^{-38}M_{\odot}
\label{MPlanck}
\end{equation} 
and using the Planck density $\rho_{Planck}$ 
\begin{equation}
\rho_{Planck} \equiv (M_{Planck})^4 \sim (10^{-5}g)(10^{-33}cm)^{-3} = 10^{94} \rho_{H_2 O}
\label{rhoPlanck} 
\end{equation}
the density of a general black hole $\rho_{BH}(M_{BH})$ is
\begin{equation}
\rho_{BH}(M_{BH}) \sim \left( \frac{M_{BH}}{r_{BH}^3} \right) = \rho_{Planck} \left(
\frac{M_{Planck}}{M_{BH}} \right)^2 \sim 10^{94} \rho_{H_2 O} \left( \frac{10^{-38} M_{\odot}}{M_{BH}} \right)^2
\label{rhoBH}
\end{equation}
which means that for a solar-mass black hole
\begin{equation}
\rho_{BH}(M_{\odot}) \sim 10^{18} \rho_{H_2 O}
\label{rhoBHSun}
\end{equation}
while for a billion solar mass black hole
\begin{equation}
\rho_{BH}(10^9M_{\odot}) \sim \rho_{H_2 O}.
\label{rhoBHSM}
\end{equation}
and above this mass the density falls as $M_{BH}^{-2}$.

\bigskip

\noindent
The mass of the Carr-Hawking PBH is derived by combining
Eqs. (\ref{rhogamma}) and (\ref{rhoBH}). We see from these 
two equations that $M_{PBH}$ grows linearly with time and using Planckian units 
or Solar units we find respectively
\begin{equation}
M_{PBH} \sim \left( \frac{t}{10^{-43} sec} \right) M_{Planck}
\sim \left( \frac{t}{1 sec} \right) 10^5 M_{\odot}
\label{PBHmass}
\end{equation}
which implies, if we perversely insisted on PBH formation
before the electroweak phase transition, $t < 10^{-12}s$, that
\begin{equation}
M_{PBH} < 10^{-7} M_{\odot}
\label{upperbound}
\end{equation}

\bigskip

\noindent
The incorrect upper bound in Eq.(\ref{upperbound}) explains historically why the MACHO
searches around 2000 \cite{MACHO,EROS}, inspired by the 1986
suggestion of Paczynski\cite{Paczynski}, lacked motivation
to pursue searching beyond $100M_{\odot}$ because it was thought 
incorrectly at that
time that PBHs were too light. It was known correctly
that the results of gravitational collapse of
normal stars, or even large early stars, were below $100M_{\odot}$.
Supermassive black holes with $M > 10^6M_{\odot}$
such as $Sag A^*$ in the Milky Way were beginning to be discovered in galactic centers
but their origin at that time was mysterious. We shall discuss this again later
in the paper.

\bigskip

\noindent
Hawking radiation implies that the lifetime for a black hole evaporating 
{\it in vacuo} is given by the cubic formula
\begin{equation}
\tau_{BH} \sim \left( \frac{M_{BH}}{M_{\odot}} \right)^3 \times 10^{64} years
\label{BHlifetime}
\end{equation}
so that to survive for the age $10^{10}$ years of the universe, there is a
lower bound on $M_{PBH}$ to augment the upper bound in Eq.(\ref{upperbound}),
giving as the full range of Carr-Hawking PBHs:
\begin{equation}
10^{-18}M_{\odot} < M_{PBH} < 10^{-7} M_{\odot}
\label{CarrHawking}
\end{equation}
The lowest mass Carr-Hawking PBH in Eq.(\ref{CarrHawking}) has the extraordinary density
$\rho \sim 10^{58} \rho_{H_2 O}$. It has the radius of a proton and
the mass of ten thousand aircraft carriers
\footnote{The radiation domination ends at $t \sim 47ky \sim 10^{12}sec$ which permits,
according to Eq.(\ref{PBHmass}), a PBH with mass $10^{17} M_{\odot}$. This
has Schwarzschild radius $\sim 10^4pc$, Hawking temperature,
according to Eq.(\ref{Hawking}), of $\sim 6 \times 10^{-25}K$,
and density, according to Eq.(\ref{rhoBH}), of $\sim 10^{-16} g/cm^3$. Such a possible
primordial super-duper-massive
black hole would be a hundred times the mass of the Virgo cluster and one
millionth the total mass of the visible universe. Such an object might be 
unassociated with any galaxy or cluster of galaxies.}.

\bigskip

\noindent
The Hawking temperature $T_H(M_{BH})$
of a black hole is
\begin{equation}
T_H(M_{BH}) = 6 \times 10^{-8}K \left( \frac{M_{\odot}}{M_{BH}} \right)
\label{Hawking}
\end{equation}
which would be above the CMB temperature, and hence there would be outgoing radiation for all of the cases with $M_{BH} < 2 \times 10^{-8}M_{\odot}$.
Hypothetically, if the dark matter halo were made entirely of the brightest possible (in terms of Hawking radiation) $10^{-18}M_{\odot}$
PBHs, the expected distance to the nearest PBH would be about $10^7$ km.
Although the PBH temperature, according to Eq. (\ref{Hawking}) is $\sim 6\times 10^{10} K$,
the inverse square law 
renders the intensity of Hawking radiation too small, by many orders of magnitude,
to allow its detection by any foreseeable apparatus on Earth.

\bigskip

\section{Parametric Resonance in Hybrid Inflation}

\noindent
The original Carr-Hawking mechanism produces PBHs with masses in the range
up to $10^{-7} M_{\odot}$. In this Section we shall exhibit formation of PBHs
by a different mechanism.
As discussed, PBH formation requires very large inhomogeneities.
Here we shall merely illustrate how to produce inhomogeneities
which are exponentially large.

\bigskip

\noindent
In a single inflation, no exceptionally large density perturbation is expected. Therefore
we use two-stage hybrid inflation with respective fields called \cite{LiddleLyth},
inflaton and waterfall. The idea of parametric
resonance is that after the first inflation mutual couplings of the
inflaton and waterfall fields cause both to oscillate wildly
and produce perturbations which grow exponentially.
The secondary (waterfall) inflation then stretches further these inhomogeneities,
enabling production of PBHs with arbitrarily high mass.
The specific model provides an existence theorem to confirm that
arbitrary mass PBHs can be produced. The resulting mass function
is spiked, but it is possible that other PBH production mechanisms can
produce a smoother mass function, as deserves further study.

\bigskip

\noindent
We follow \cite{FKTY} in using a supergravity framework, defining
by $S$ the inflaton superfield and by $\Psi, \bar{\Psi}$ the waterfall superfields.
The superpotential is
\begin{equation}
W = S \left( \mu^2 + \frac{(\bar{\Psi}\Psi)^2}{M^2} \right)
\label{superpot}
\end{equation}
in which $\mu$ is the inflation scale and $M$ is a cut-off.

\bigskip

\noindent
The Kahler potential is
\begin{equation}
K = |S|^2 + |\Psi|^2 + |\bar{\Psi}|^2
\label{Kahler}
\end{equation}
and from Eqs.(\ref{superpot}) and (\ref{Kahler}) the potential is
\begin{equation}
V(\sigma,\psi) \sim \left(1 + \frac{\sigma^4}{8} + \frac{\psi^2}{2} \right)
\left( -\mu^2 + \frac{\psi^4}{4 M^2} \right)^2 + \frac{\sigma^2 \psi^6}{16 M^4}
\label{potential}
\end{equation}
where we have defined $\psi = 2 \Re (\Psi)$ and $\sigma = \sqrt{2} \Re (S)$
with $\Re \equiv$ real part.

\bigskip

\noindent
Stationarizing Eq.(\ref{potential}) gives vacua at $\sigma=0$ and $\psi=2\sqrt{\mu M}$.
For the case $\sigma > \sqrt{\mu M}/2$
there is a $\sigma$-dependent minimum for $\psi$ at
\begin{equation}
\psi_0 \sim \left(\frac{2}{\sqrt{3}} \right) \left( \frac{\mu M}{\sigma} \right).
\label{psimin}
\end{equation}

\bigskip

\noindent
Because $\psi$ has a large mass, it rolls to $\psi_0$ and integrating it out
results in the potential
\begin{equation}
V(\sigma) = \mu^4 \left( 1 + \frac{\sigma^4}{8} - \frac{2}{27}\frac{\mu^2M^2}{\sigma^4} \right)
= \mu^4 + \frac{\mu^4}{8} \left( \sigma^4 - \sigma_0^4 \left(\frac{\sigma_0}{\sigma} \right)^4
\right )
\label{sigmapot}
\end{equation}
in which $\sigma_0 = \sqrt{2}/3^{\frac{3}{8}} (\mu M)^{\frac{1}{4}}$. So long as the first term
in Eq.(\ref{sigmapot}) is largest, the inflaton slow rolls.

\bigskip

\noindent
After this inflation, the $\sigma$ and $\psi$ fields oscillate, decaying into their quanta via their self and mutual couplings. Specific modes of $\sigma$ and $\psi$ are amplified by
parametric resonance.

\bigskip

\noindent
From Eq.(\ref{sigmapot}), we may write the equation of motion for a Fourier mode
$\sigma_k$ as
\begin{equation}
\sigma_k^{''} + 3H \sigma_k^{'} +
\left[ \frac{k^2}{a^2} + m_{\sigma}^2 + 3m_{\sigma}^2 \frac{\tilde{\psi}}{\sqrt{\mu M}}
\cos(m_{\sigma}t) \right] \sigma_k \sim 0
\label{Mathieu}
\end{equation}
where we defined $m_{\sigma}= \sqrt{8\mu^3/M}$. and $\tilde{\psi}$ is the 
amplitude of $\psi$ oscillations.

\bigskip

\noindent
Eq.(\ref{Mathieu}) is recognized to be of Mathieu type 
with the required exponentially-growing solutions. Numerical
solution shows that the peak wave number $k_{peak}$ is approximately linear in
$m_{\sigma}$. The resultant PBH mass, the horizon 
mass when the fluctuations re-enter the horizon, is approximately
\begin{equation}
M_{PBH} \sim 1.4 \times 10^{13}M_{\odot} \left( \frac{k_{peak}}{Mpc^{-1}} \right)^{-2}
\label{PBHmass}
\end{equation}

\bigskip

\noindent
Explicit plots were exhibited in \cite{FKTY} 
for the cases $M_{PBH} = 10^{-8}M_{\odot},
 10^{-7}M_{\odot}$ and $10^5M_{\odot}$
but it was checked that the parameters 
can be chosen to produce arbitrary
PBH mass. 

\bigskip

\noindent
In this production mechanism based on hybrid inflation with parametric 
resonance, the mass function is sharply spiked at a specific mass region.
Whether such a mass function is a general feature of PBH formation, or
is only a property of this specific mechanism, merits further study.

\bigskip

\section{Dark Matter and Supermassive Black Holes}

\bigskip

\noindent
In Section 2 we discussed the
method of producing PBHs proposed by
Carr and Hawking.
Insisting that the production take place
before the electroweak phase transition,
and bearing in mind the survival to the age
of the universe from Hawking radiation,
led us to a range of possible PBH masses
from $10^{-18}M_{\odot}$
to $10^{-7}M_{\odot}$.

\bigskip

\noindent
In Section 3, using a different production mechanism
based on parametric resonance in hybrid inflation
this was augmented to a much bigger mass range
\begin{equation}
10^{-18} M_{\odot} < M_{PBH} <  10^{17}M_{\odot}
\label{PBHrange}
\end{equation}
which adds to Carr-Hawking, {\it inter alia},
Primordial Intermediate-Mass Black Holes (PIMBHs)
in the range
\begin{equation}
10^2 M_{\odot} < M_{PIMBH} < 10^6 M_{\odot}
\label{PIMBH}
\end{equation}
and Primordial Supermassive Black Holes (PSMBHs)
in the range
\begin{equation}
10^6 M_{\odot} < M_{PSMBH} < 10^{17} M_{\odot}.
\label{PIMBH}
\end{equation}
where we have truncated the upper end at $10^{17}M_{\odot}$ as the heaviest
conceivable black hole likely to exist in the Universe.

\bigskip

\noindent
For dark matter in galaxies, PIMBHs are important,
 where the upper end may be truncated at 
$10^{5}M_{\odot}$ to stay well away 
from galactic disk instability \cite{Ostriker}. 
For supermassive black holes in galactic cores,
PSMBHs are natural candidates, as they are also for 
a part of the dark matter in clusters.

\subsection{Dark Matter in Galaxies}

\noindent
The dark matter in the Milky Way fills out an approximately spherical halo
somewhat larger in radius than the disk occupied by the luminous stars.
Numerical simulations of structure formation suggest a profile of the
dark matter of the NFW types\cite{NFW}. The NFW profile is 
fully independent
of the mass of the dark matter constituent.

\bigskip

\noindent
Our discussion\cite{Frampton} focused on galaxies like the Milky Way
and restricted the mass range for the appropriate dark matter 
to only three orders of magnitude
\begin{equation}
10^2 M_{\odot} < M < 10^5M_{\odot}
\label{Frampton}
\end{equation}

\bigskip

\noindent
We shall not repeat the arguments here, just to say that the
constituents are Primodial Intermadiate Mass Black Holes, PIMBHs. Given a
total dark halo mass of $10^{12}M_{\odot}$, the number $N$ of PIMBHs
is between ten million ($10^7$) and ten billion ($10^{10}$) 
Assuming the dark halo has radius $R$ of a hundred thousand ($10^5$)
light years
the mean separation $\bar{L}$ of PIMBHs can be estimated by
\begin{equation}
\bar{L} \sim \left( \frac{R}{N} \right)
\label{meanL}
\end{equation}
which translates to
\begin{equation}
100 ly < \bar{L} < 1000 ly
\label{meanL2}
\end{equation}
which is also an estimate of the distance of the nearest PIMBH
to the Earth.

\bigskip

\noindent
It may be surprising that as many as $10^7 \leq N \leq 10^{10}$ intermediate-mass black holes
in the Milky Way have remained undetected. They could have been
detected more than a decade ago had the MACHO Collaboration
\cite{MACHO} persisted in its microlensing experiment at Mount Stromlo
Observatory in Australia. We shall return to this point in our final discussion.

\bigskip

\subsection{Dark Matter in Clusters}

\noindent
The first discovery of dark matter by Zwicky \cite{Zwicky,Zwicky2} 
was in the Coma cluster which is a large cluster at 99 Mpc
containing over a thousand galaxies and with
total mass estimated at $6 \times 10^{14}M_{\odot}$ \cite{Merritt}.
A nearer cluster at 16.5 Mpc is the Virgo cluster with over two thousand
galaxies and whose mass $\sim 10^{15}M_{\odot}$
is also dominated by dark matter, as well as a small
amount of X-ray emitting gas \cite{Virgo1,Virgo2}.
A proof of the existence (if more were needed)
of cluster dark matter was provided by the Bullet cluster
collision where the distinct behaviors of the X-ray emitting gas which
collides, and the dark matter which does not collide, was clearly observable
\cite{Bullet}.

\bigskip

\noindent
Since there is not the same disk stability limit as for galaxies,
the constituents of the cluster dark matter can involve also PSMBHs 
up to much higher masses.
In the Universe, we may speculate here that there may be unassociated PBHs 
with any mass
up to $10^{17}M_{\odot}$ drifting outside of any galaxy or cluster
of galaxies.

\bigskip

\subsection{Supermassive Black Holes at Galactic Centres}

\noindent
As mentioned in the Introduction, in the Milky Way there is 
SMBH, $Sag A*$, with mass
$M_{Sag A*} \sim 4 \times 10^6 M_{\odot}$. Other galaxies have
SMBHs with masses ranging up to $2.1 \times 10^9M_{\odot}$
(for the galaxy NGC4889). Only a tiny fraction of 
galaxies have been studied, so the range
of galaxies' core SMBHs is likely broader.

\bigskip

\noindent
A black hole with the mass of $Sag A^*$ would disrupt the disk
dynamics \cite{Ostriker} were it out in the spiral arms but at, or
near to, the
center of mass it is more stable. $Sag A^*$ is far too massive to
have been the result of a gravitational collapse, and if we take
the view that all black holes either are 
the result of gravitational collapse or are primordial then the
galaxies' core SMBHs must be primordial. This offers
a new explanation of their origin.

\bigskip

\subsection{Galaxy formation}

\noindent
If our discussion is correct, it provides a clear time-ordering for galaxy formation
that the dark matter precedes star formation by half a billion years.
Let us consider the history of the Milky Way.

\bigskip

\noindent
The constituents of the Milky Way's dark matter halo, PIMBHs,
were produced in the era of radiation domination which ended at
time $t \sim 47 ky$ (red shift $Z \sim 4760$). 
Only much later, after 560 million years ($Z\sim 8$),
did star formation begin in the Milky Way.

\bigskip

\noindent
In this version of cosmic history, 
much of the large-scale structure
formation including of galaxies such as the Milky Way
progresses during the half billion years
represented by the red shifts $4760 > Z > 8$. 
This stage importantly involves {\it only} dark matter.
Baryonic astrophysical objects like the Solar System
appear only when $Z < 8$ and 
are demoted to an afterthought with respect to
the Milky Way's formation.

\bigskip

\section{Discussion}

\bigskip

\noindent
Such a bold solution of the dark matter problem cries out for experimental
verification. Three methods have been discussed: wide binaries,
distortion of the CMB, and microlensing. Of these, microlensing seems the most
direct and the most promising.

\bigskip

\noindent
Microlensing experiments were carried out by the MACHO\cite{MACHO}
and EROS\cite{EROS} Collaborations several years ago. At that time, it was
believed that PBH masses were below $10^{-7}M_{\odot}$ by virtue of the
Carr-Hawking mechanism. Heavier black holes could, it was then believed,
arise only from gravitational collapse of normal stars, or 
heavier early stars, and would have mass below $100M_{\odot}$. 

\bigskip

\noindent
For this reason, there was no motivation to suspect that there might be
MACHOs which led to higher-longevity microlensing events.
The longevity, $\hat{t}$, of an event is
\begin{equation}
\hat{t} = 0.2 yrs \left( \frac{M_{PBH}}{M_{\odot}} \right)^{\frac{1}{2}}
\label{longevity}
\end{equation}
which assumes a transit velocity $200 km/s$.
Subsituting our extended PBH masses, one finds approximately
$\hat{t} \sim 6, 20, 60$ years for $M_{PBH} \sim 10^3, 10^4, 10^5 M_{\odot}$
respectively, and searching for light curves with these higher
values of $\hat{t}$ could be very rewarding.

\bigskip

\noindent
Our understanding is that the original telescope used by the
MACHO Collaboration\cite{MACHO} at the Mount Stromlo
Observatory in Australia was accidentally
destroyed by fire, and that some
other appropriate telescopes are presently being used to search
for extasolar planets, of which two thousand are already known. 

\bigskip

\noindent
It is seriously hoped that MACHO searches will resume and
focus on greater longevity microlensing events. Some encouragement
can be derived from this, written this month by a member of the 
original MACHO Collaboration :

\bigskip

\noindent
{\it There is no known problem with searching for events of greater longevity than those discovered in 2000; only the longevity of the people!}

\bigskip

\noindent
That being written, convincing observations showing only a fraction of the light curves
could suffice? If so, only a fraction of the {\it e.g.} six years, corresponding to PIMBHs with
one thousand solar masses, could well be enough to confirm the theory.

\bigskip

\noindent
Finally, going back to the 2010 Vera Rubin quote mentioned in the Introduction, it is

\bigskip

\noindent
{\it "If I could have my pick, I would like to learn that Newton's laws must be modified in order to correctly describe gravitational interactions at large distances. That's more appealing than a universe filled with a new kind of sub-nuclear particle."}

\bigskip

\noindent
If our solution
for the dark matter problem is correct, Rubin's preference for no new elementary particle
filling the Universe would be vindicated, because for dark matter microscopic particles become irrelevant.
Regarding Newton's law of gravity, it would not need modification beyond general relativity theory which is needed for the black holes. In this sense, Rubin did not need to pick either alternative to explain dark matter.

\end{document}